\documentclass[%
 reprint,
 amsmath,amssymb,
 aps,showkeys
]{revtex4-2}
\usepackage{graphicx}
\usepackage{dcolumn}
\usepackage{bm}

\begin{document}

\title{The chiral beat of algal flagella: force and torque via imaging}
\author{Laurence G. Wilson}
 \email{laurence.wilson@york.ac.uk}
\affiliation{School of Physics, Engineering and Technology, University of York, Heslington, York, YO10 5DD, United Kingdom}
 \author{Martin A. Bees}%
 \email{martin.bees@york.ac.uk}
\affiliation{Department of Mathematics, University of York, Heslington, York, YO10 5DD, United Kingdom}%

\date{\today}

\begin{abstract}
Flagella allow eukaryotic cells to move and pump fluid.  We present the first three-dimensional, time-resolved imaging of the flagellar waveform of {\it Chlamydomonas reinhardtii}, a model alga found in fresh water.  During the power stroke, we find that the flagella show rotational symmetry about the cell's centreline, but during the recovery stroke they display mirror  symmetry about the same axis.  We use our three-dimensional imaging data to test the applicability of resistive force theory when a force-free configuration is approximated using a cell on a micropipette.  The inferred values of cells' swimming speeds and rotation rates show good agreement with experimental results on freely swimming cells.  
\end{abstract}

\maketitle

Symmetry and symmetry breaking are central concepts in physics that offer a powerful perspective on phenomena in the life sciences.  Many biological structures are highly symmetrical in structure but have slight modifications that serve to break symmetry and fine-tune their function.  One such structure is the eukaryotic flagellum, an organelle that is highly conserved across the eukaryotic domain \cite{KhanJS_CurrBiolRev18}.  The flagellum's central component, the axoneme, has a recognizable structural motif:  a chiral arrangement of nine microtubule doublets encircling a central pair of microtuble singlets.  The doublets slide length-wise relative to each other under the action of dynein molecular motors, resulting in the whip-like motion from which the flagellum gets its name.  The function of this complex molecular machine has been the focus of study for around a century,  not least due to the axoneme's role in human reproduction:  flagella are responsible for propelling sperm cells, and pumping fluid in the the fallopian tubes.  An aspect of the movement of flagella that has hitherto been difficult to access is the three-dimensional shape of the flagellar beat.  This is challenging to study due to the small size and rapid motion of most flagella --- although they vary in length considerably,  their diameter is typically less than 500\, nm and they beat at 10--100\, Hz.   The green alga {\it Chlamydomonas reinhardtii} is a model species for the study of flagella; its genetics are relatively well-known, and a wide range of mutants have been created.  The cells naturally occur in open freshwater and swim with a `breaststroke' motion, pulled by a pair of anterior flagella.  Ultrastructural studies of {\it C. reinhardtii} flagella have observed the internal configuration of the axonemes and dynein motors with precision  \cite{HoopsGW_JCellBiol83,NicastroCSJPRGMPJM_Science06,BuiRYRKTI_JCellBiol12}.  Careful studies of mutant {\it Chlamydomonas} strains have shown that when structures that ensure the symmetry of the flagllar orientations are disrupted, flagellar function is impaired and the cells' ability to swim is degraded \cite{HoopsRWJJGW_JCellBiol84}.  Studying the motion of flagella in freely-swimming cells is challenging because {\it C. reinhardtii} swim at around 10 body lengths per second, and rotate as they do so, occluding the view of the flagella.  To mitigate this, glass micropipettes have been used to hold and position individual cells \cite{HoopsRWJJGW_JCellBiol84,RufferWN_CellMoti90,RufferWN_CellMotil91,PolinITKDJGRG_Science09}, allowing the flagellar dynamics of these fast swimmers to be studied in detail.

\begin{figure*}[ht]
\includegraphics[width=1\linewidth]{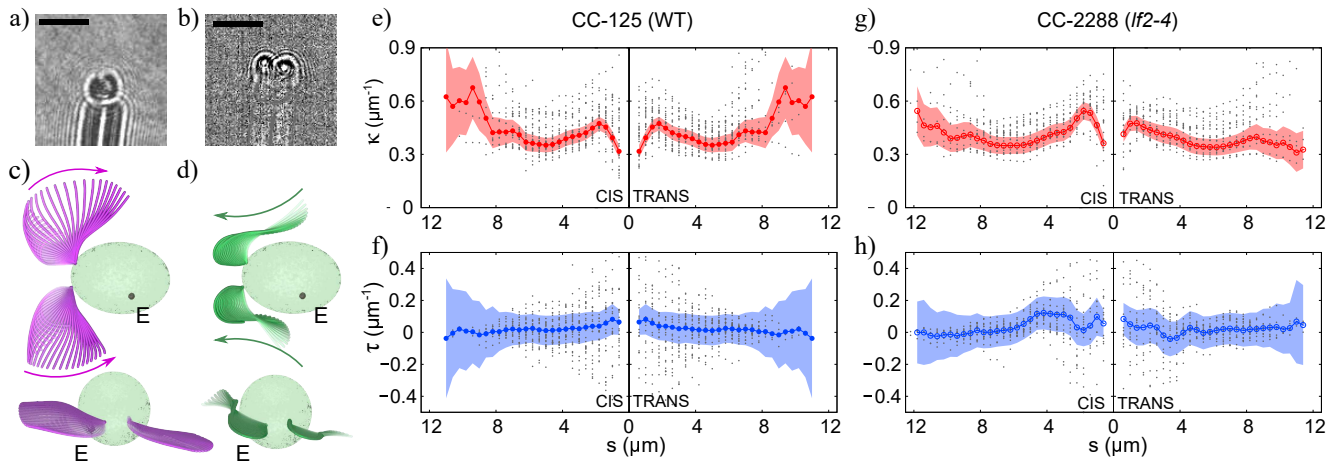}
\caption{\label{fig:Fig1}(a) Raw holographic data showing an algal cell on the end of a micropipette, approximately 5\,$\mu$m below the focal plane (scale bar shows 10\,$\mu$m).  (b) Holographic data after application of a Kalman filter, showing the diffraction patterns from flagella. (c),(d)  Rendering of a reconstructed cell body and flagella positions from holographic data during power and recovery stroke, respectively. (e),(f), Mean curvature and torsion of the cis and trans flagella as a function of length, for the wildtype strain CC-125.  The gray data points correspond to values from individual movies, and the shaded areas indicate a 95\% CI.  (g),(h) Mean curvature and torsion values for cells from the mutant strain CC-2288 that exhibits slightly longer flagella than the wildtype.}
\end{figure*}

Previous studies that have approached the problem of measuring the flagellar waveform have used extrapolation based on two-dimensional images \cite{CorteseKW_PRL21}, or multi-plane imaging of flagella that had been isolated from the cell body,  demembranated and chemically reactivated \cite{MojiriSISMHJABIGAGJE_BiomedOptExp21}.  We use digital holographic microscopy (DHM) to study intact {\it C. reinhardtii} cells, to measure the shape of the flagellar beat, and evaluate the applicability of a classic low Reynolds number hydrodynamics approximation to this species.  We use DHM in an `inline' configuration, which can be implemented easily on a commercial microscope \cite{Kim_SPIERev10}.  DHM has been used to track swimming microorganisms including bacteria \cite{CheongCWYGMNYCSPLKCL_BiophysJ15,MolaeiMBRSJS_PRL14,BianchiFSRD_PRX17,KuhnFSNFFRBHLWBEKT_NatCommun18,GibsonMBESCLJN_JOSAA21}, archaea \cite{ThorntonJBSDBBLW_NatComms20}, dinoflagellates \cite{ShengEMJKJARBAP_PNAS07}, sperm cells \cite{SuLXAO_PNAS12,JikeliLABFLWRPRCMPARCBUK_NatComm15,GongSRGGUKJEBFLA_EPJE21}, and other eukaryotes \cite{WilsonLCSR_PNAS13,FindlayMOKSPKPWLW_eLife21}.  In the inline configuration, a sample chamber is illuminated with coherent light, part of which is scattered by the objects of interest.  This scattered light interferes with the unscattered light at the image plane, yielding images such as the one depicted in Fig. \ref{fig:Fig1}a.  The flagella of {\it C. reinhardtii} are challenging to image because of their proximity to the highly-scattering cell body.  However, after careful background subtraction (see methods), the dynamic parts of the image are revealed, as shown in Fig. \ref{fig:Fig1}b.  These images are processed numerically using Rayleigh-Sommerfeld back propagation \cite{LeeDG_OE07} and an object detection scheme based on the Gouy phase anomaly \cite{WilsonRZ_OptExp12} to localise the flagellar contour in three dimensions.  We analyse the shape of the flagella using the Frenet-Serret formulae to describe a curve in three dimensions in terms of unit tangent ($\hat{\mathbf{t}}$), normal ($\hat{\mathbf{n}}$) and binormal ($\hat{\mathbf{b}}$) vectors, related by:

\begin{equation*}
\frac{\textrm{d}}{\textrm{d}s}\hat{\mathbf{t}}=\kappa\hat{\mathbf{n}} \quad\quad\quad \frac{\textrm{d}}{\textrm{d}s}\hat{\mathbf{n}}=-\kappa\hat{\mathbf{t}}+\tau\hat{\mathbf{b}} \quad\quad\quad \frac{\textrm{d}}{\textrm{d}s}\hat{\mathbf{b}}=-\tau\hat{\mathbf{n}},
\end{equation*}
where $s$ is arclength, $\kappa$ is curvature and $\tau$ is torsion.   

We studied two strains of {\it C. reinhardtii}:  CC-125 (wildtype) and CC-2288 (a `long flagella' mutant).  Figures \ref{fig:Fig1}c,d show the shape of the flagella during the power and recovery stroke, respectively.  Figures \ref{fig:Fig1}e,f show the cell- and time-averaged flagellar curvature and torsion for CC-125 cells (N=20 cells) and Figs. \ref{fig:Fig1}g,h show the equivalent data for CC-2288 cells (N=16 cells).  In both strains, the curvature and torsion are well resolved for low values of the contour length $s$ (i.e. close to the cell body) but become noisy at the distal ends of the flagella due to poor statistics (i.e. few cells have the longest flagella, and there are occasional `missed detections' due to camera noise).  Both the wild type and mutant strains show modest average torsion, with zero average torsion lying within the 95\% confidence intervals.  However, it should be noted that for straight sections of the flagellum, the curvature is zero, in which case the torsion is undefined.  Consequently, when the flagellum is approximately straight, noise in the estimated position can lead to relatively large fluctuations in torsion.  Below, we examine  the torsion during instances of high curvature when the quantities are well defined  (Fig. \ref{fig:Fig3} and accompanying text).  

\begin{figure}[ht]
\includegraphics[width=0.7\linewidth]{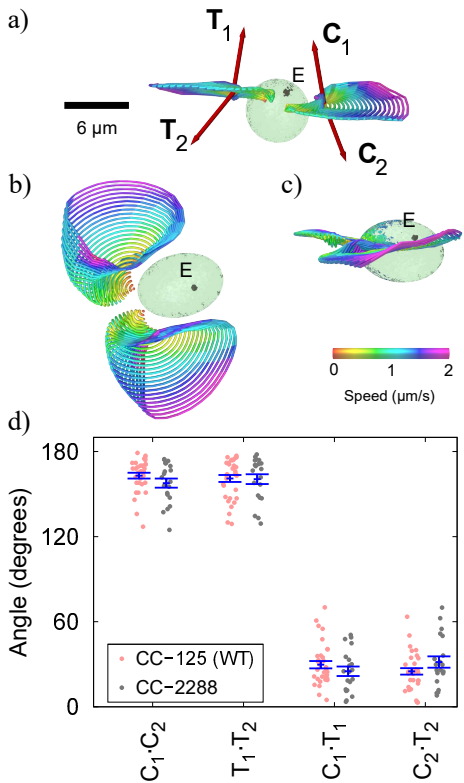}
\caption{\label{fig:Fig2}(a) Illustration of beat plane normal vectors for the power and recovery stroke of wildtype {\it C. reinhardtii}, for the cis ($\mathbf{C}_{1,2}$) and trans ($\mathbf{T}_{1,2}$) flagella.  {\bf E} indicates the position of the eye-spot.  (b) Top view of the flagellar beat plane.  (c) side view of the flagellar beat plane, illustrating the offset flagellar planes.  (d) Angles between power and recovery stroke beat planes, for strains CC-125 and CC-2288.  The blue data points represent the mean angles between the sets of normal vectors, and the error bars on these represent standard error on the mean.}
\end{figure}

\begin{figure*}[ht]
\includegraphics[width=1\linewidth]{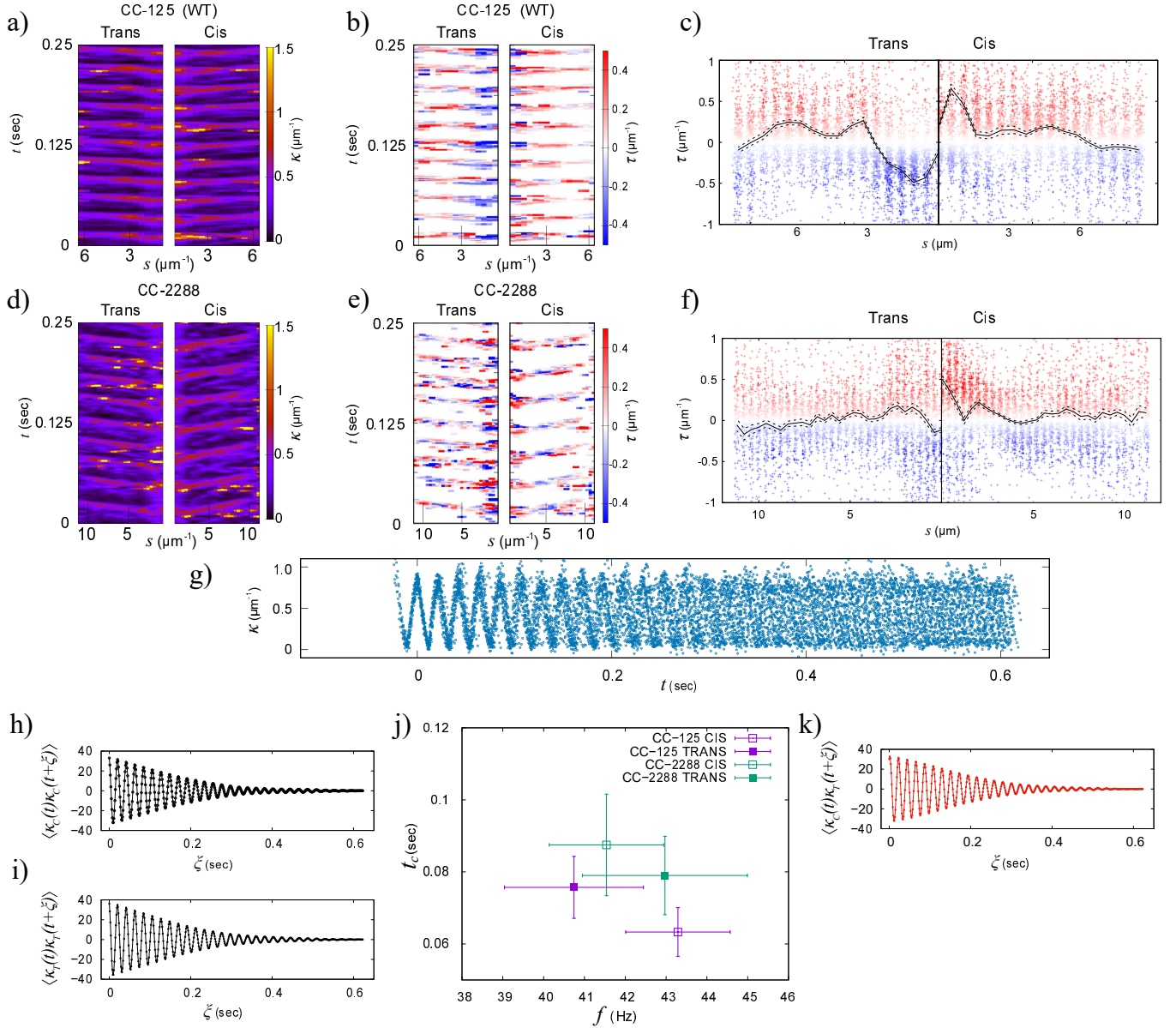}
\caption{\label{fig:Fig3} (a) Kymograph showing flagellar curvature during 22 beat cycles, as a function of contour length ($s$) and time ($t$) for wildtype strain CC-125.  Waves of curvature are initiated at the base of the flagellum and propagate to the tip in both cis and trans cases. (b) Torsion in each flagellum, during the highest-curvature portion of the beat cycle (i.e. when torsion is clearly defined).  Note the predominance of negative torsion in the trans flagellum (blue colors) and positive torsion in the cis flagellum (red colors) near the flagellar base.  (c) Average and instantaneous torsion during the high-curvature recovery stroke over approximately 600 beat cycles (12 seconds), highlighting the average negative (positive) torsion in the proximal region of the trans (cis) flagellum.  The black line denotes the mean, and the dashed lines represent standard error on the mean.  (e),(f) flagellar torsion in the recovery stroke of `long flagella' mutant strain CC-2288.  The same qualitative trend in torsion around the flagellar base is observed, albeit with a smaller magnitude.  (g)  Loss of coherence in flagellar beating in wildtype strain CC-125.  $\kappa$ at $s=2$\,$\mu$m was examined over the course of a 12 second movie, divided into 0.6 second sections and aligned according to the first maximum in curvature.  The waves of curvature lose coherence at approximately 0.6 seconds.  (h)  Autocorrelation of curvature in the cis flagellum at a contour length of $s=2$\,$\mu$m, as a function of delay time $\xi$.  The coherence time $t_c$ was extracted by fitting these data with an exponentially damped cosine function (see text).  (i) Equivalent data to that in panel (h), but from the trans flagellum.  (j)  Coherence time $t_c$ as a function of beat frequency $f$ in the cis and trans flagella of strains CC-125 and CC-2288).  The error bars denote standard error on the mean.  (k) Cross-correlation of curvature in cis and trans flagellum from a cell of wildtype strain CC-125.  The decorrelation occurs slightly faster than in panels (h), (i) due to occasional period of asymmetric flagellar beating before synchronisation is re-established.}
\end{figure*}

To examine the overall geometry of the flagellar beat envelope, we average over 500 beat cycles and record the trajectory of points at a constant arclength $s$, as shown in Fig. \ref{fig:Fig2}a,b,c.  The surface normal vectors of the planes traced out by the power and recovery stroke are indicated by their corresponding vectors $\mathbf{C_1},\mathbf{T_1}$ (power stroke) and $\mathbf{C_2},\mathbf{T_2}$ (recovery stroke), where $\mathbf{C}$ and $\mathbf{T}$ vectors indicate cis and trans respectively.   Without loss of generality, we set the power stroke for the cis flagellum, ${\bf C}_1$, to point upwards in our coordinate system, and define the other vectors ($\mathbf{C_2},\mathbf{T_1},\mathbf{T_2}$) from this.  When averaged across all of the cells in our study, we find that the angle between power and recovery stroke planes is about 160 degrees in both CC-125 and CC-2288.  The angle between the cis and trans flagellar beat planes is around 30 degrees, in good agreement with previous work that estimated the beat plane orientation based on a 2D projection \cite{CorteseKW_PRL21}.  Unlike previous studies, we also find that normal vectors to the power and recovery stroke beat planes are not antiparallel:  $\mathbf{C_2} \cdot \mathbf{T_2} < \mathbf{C_1} \cdot \mathbf{T_1}$.  This asymmetry between power and recovery strokes illustrates a subtlety in the mechanical coupling between the flagella and the basal apparatus; if the flagellar beat envelopes were simply rotated copies of each other, we would find $\mathbf{C_1} \cdot \mathbf{T_1} = \mathbf{C_2} \cdot \mathbf{T_2}$.  Figure \ref{fig:Fig2}d shows results from all of the cells in our study.  The average beat plane orientations are the same across both strains (CC-125 and CC-2288), consistent with the mutant cell type having flagella that are longer but otherwise similar to wildtype.

When examined in closer detail, intriguing differences emerge between the cis and trans flagellar waveforms during the power and recovery strokes.  At the start of the power stroke, both flagella demonstrate a small `shrugging' motion close to the base of the flagella (see Supplementary Movie S1).  This motion is reminiscent of the flagellum of {\it Euglena}, a monoflagellate alga \cite{RossiGCABGNAD_PNAS17}.  Both cis- and trans flagellar waveforms show the same chirality during this phase of the beat --- there is rotational symmetry about the cell's centerline.  During the power strokes, the flagella are at their straightest, while they are most curved during the recovery stroke.  Intriguingly, unlike the power stroke, the recovery stroke appears to show mirror symmetry about the cell centerline.

We have created a kymograph of the curvature as a function of time for both wildtype and mutant strains, shown in Fig. \ref{fig:Fig3}a,d.  The light colored bands corresponding to high curvature propagate from the proximal to the distal ends of the flagella at speeds around 1\,mm/s.  The motion repeats with a frequency of 40--50\,Hz.  The waves of high curvature correspond to the \emph{recovery stroke} of the beat, as the flagella move to the front of the cell.  Figure \ref{fig:Fig3}b,e show the corresponding waves of torsion that propagate with the highly curved sections.  We have applied a cut-off to the data so that torsion is only plotted when the local curvature exceeds $\sim 0.4 \mu$m$^{-1}$, when torsion is relatively well defined.  A notable feature of the torsion in both the wildtype and mutant cells is the opposite handedness of the proximal section at the start of the recovery stroke: the trans flagellum has predominantly negative torsion, while the cis flagellum has predominantly positive torsion.  Figure \ref{fig:Fig3}c,f show the average torsion over approximately 600 beat cycles for the wildtype and mutant cell lines, during the recovery stroke only.  The opposite polarity of the torsion can be seen most clearly in the wildtype (Fig. \ref{fig:Fig3}c), although the same trend is present and statistically significant in CC-2288.  These subtle differences in the symmetries displayed by the flagella are likely to be a consequence of the interactions between inherent structural chirality (dynein molecules are arranged in chiral fashion between the axonemal doublets), physical coupling to the basal body apparatus and hydrodynamic coupling to the cell body.  The interaction is complicated, and shows the versatility of the axoneme and flagellum in being able to generate qualitatively different waveforms depending on external constraints.

Next, we investigated the time-dependent nature of the flagellar beat and its coherence.  The synchronization of multiple flagella on a single cell body is the subject of previous work focusing on organisms with one, two or many flagella (cilia)  \cite{BrumleyMPTPRG_PRL12,GeyerFJJHBF_PNAS13,BrumleyKWMPRG_eLife14,KlindtCRCWBF_NewJPhys17}.  Discrete phase slips have been observed to reset the beat of {\it C. reinhardtii} when it becomes desynchronized \cite{PolinITKDJGRG_Science09}, at least when cells are restrained by a micropipette.  We have chosen to focus on the short time coherence of the flagellar beat, that is, how the beat frequency (and hence phase) fluctuates stochastically over time.  Figure \ref{fig:Fig3}g shows the curvature of a section of a wildtype flagellum, taken from Fig. \ref{fig:Fig3}a at $s=$2\,$\mu$m from the cell body.  The time series has been divided up in to 0.6 second `blocks' and aligned so that the first peak in each series overlaps.  The curvature is periodic across all instances, but the beat frequency drifts enough to smear out the sinusoidal curvature after about 0.5 seconds.  To further quantify this, we calculated the autocorrelation of the curvature in the cis and trans flagella, $\kappa_C$ and $\kappa_T$ respectively, and have plotted these quantities in Fig. \ref{fig:Fig3}h,i.  The cross-correlation of curvature between the two flagella is shown in Fig. \ref{fig:Fig3}k.  In all three cases, the correlation functions take the form of a damped cosine function, with a characteristic decay time $t_c = 0.08$ sec, corresponding to 3--4 flagellar beat cycles.  Figure \ref{fig:Fig3} summarises the characteristic decay time of these autocorrelation functions across all cells in the data set as a function of beat frequency; we do not find a strong dependence of beat coherence on this quantity.

\clearpage
\begin{figure}[ht]
\includegraphics[width=0.7\linewidth]{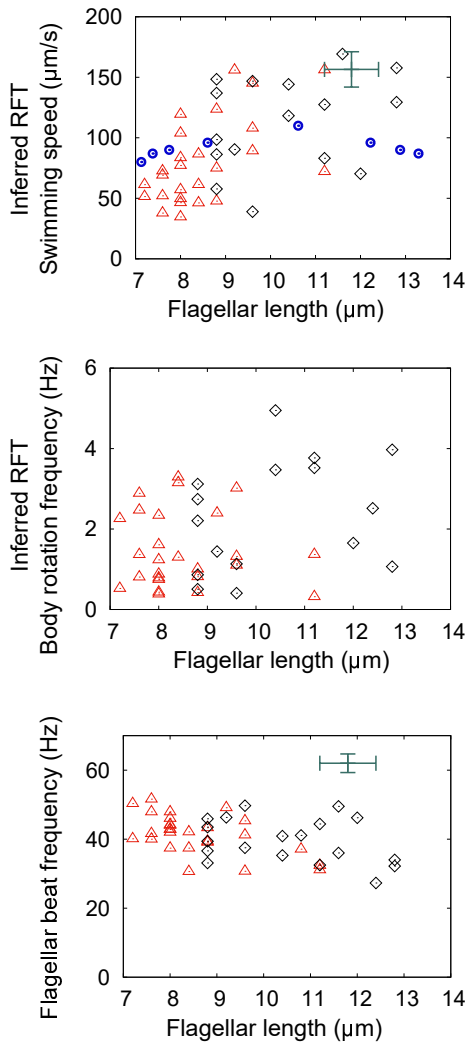}
\caption{\label{fig:Fig4}Results of a resistive force theory analysis of three-dimensional imaging data.  Each data point represents one cell of wildtype strain CC-125 (red triangles, 21 cells) or `long flagella' mutant CC-2288 (black diamonds, 19 cells). Experimental data from freely swimming cells taken from references \cite{KhonaVRMMPVAKKDSSLBJDADSMDMJD_JBiolPhys13} (green crosses, error bars represent standard deviation) and \cite{BauerHIKWNHJKWM_iScience21} (blue circles). (a)  Predicted swimming speed as a function of total flagellar length of a cell.  These values are obtained by determining the thrust due to the flagella and calculating the resultant speed of the cell body  (see text).  (b) Predicted rotation speed of the cell body about the swimming direction, owing to the rotational offset and torsion of the flagella.  (c)  Average flagellar beat frequency obtained from a Fourier analysis of flagellar curvature, $\kappa$(t).}
\end{figure}

Finally, we use our three-dimensional coordinates to test the applicability of a long-standing approximation of flagellar hydrodynamics:  resistive force theory (RFT).  This approximation can be used to model the force ($\mathbf{F}$) and torque ($\mathbf{\Gamma}$) on the cell body due to the motion of its swimming apparatus.  In RFT, a flagellum is typically modeled as locally straight, thin rod with drag coefficients that are different for motion tangential ($K_{\hat{t}}$) and normal ($K_{\hat{n}}$) to the rod axis.  Values for these coefficients were obtained by Lighthill \cite{Lighthill_SIAMRev76}, in an analytical study of flagellar fluid dynamics.  The net forces and torques on a freely swimming cell must equal zero in the absence of external constraints:
\begin{align}
   \mathbf{F}_b + \Sigma_i \delta \mathbf{F}_i &= \mathbf{0} \label{eqn:Fbalance}\\
   \mathbf{\Gamma}_b +  \Sigma_i \left( \mathbf{r}_i\wedge\delta\mathbf{F}_i \right)+ \Sigma_i\mathbf{\Gamma}_i &= \mathbf{0} \label{eqn:Gbalance},
   \end{align}
where the subscript $b$ refers to the body, $\delta\mathbf{\Gamma}_i$ is an element of torque caused by the rotation of a segment about its hydrodynamic centre, and $\delta\mathbf{F}_i$ is an element of force caused by the motion of a rod-like segment at position $\mathbf{r}_i$ relative to the hydrodynamic centre of the system.  In our experiments, the cell experiences an external force from the micropipette, and so the flow fields generated by the cell body and flagella will be qualitatively different from those produced by a freely-swimming individual.   In order to improve this approximation, we transform from the body-centred frame (indicated by primed quantities) in which we perform our imaging, to a notional lab frame in which the cell would swim freely.  The hydrodynamic force $\delta\mathbf{F}$ experienced by a segment of the flagellum at position $s$ and time $t$ is resolved into components tangential and normal to the segment \footnote{To clarify, as we are modeling the flagellar element as a straight rod, the binormal direction is not defined.  The `normal' component here refers to all non-tangential forces.}.  Hydrodynamic coupling with the cell body and between neighbouring segments is ignored.  For a freely-swimming cell with velocity $\mathbf{v}_b$, the force experienced by a flagellar element moving with velocity $\mathbf{v}_i=\mathbf{v}'_i-\mathbf{v}_b$ in the lab frame is given by:
\begin{equation}
\delta\mathbf{F} = \textrm{d}s \left[K_{\hat{t}}((\mathbf{v}'-\mathbf{v}_b)\cdot\mathbf{\hat{t}})\mathbf{\hat{t}} + K_{\hat{n}}((\mathbf{v}'-\mathbf{v}_b)\cdot\mathbf{\hat{n}})\mathbf{\hat{n}} \right].
\end{equation}
The force and torque acting on the cell body are given by 
\begin{align}
\mathbf{F}_b &= 6\pi\eta a \mathbf{v}_b \\
\mathbf{\Gamma}_b &= 8\pi\eta a^3 \mathbf{\Omega}_b,
\end{align} 
where $\eta$ is the dynamic viscosity of the medium, $a$ is the cell body radius (obtained for each cell from the microscopy data), and $\mathbf{\Omega}_b$ is the angular velocity of the cell body.  By solving Eqs. \ref{eqn:Fbalance} and \ref{eqn:Gbalance}, we obtain estimates of the swimming speed and rotation frequency of the cells.  These data, along with a measurement of flagellar beat frequency are plotted in Fig. \ref{fig:Fig4} as a function of flagellar length.  We compare the predictions produced by the RFT model to experimental data presented in two experimental studies of freely-swimming cells with flagella of varying length \cite{KhonaVRMMPVAKKDSSLBJDADSMDMJD_JBiolPhys13,BauerHIKWNHJKWM_iScience21}.  We find that the swimming speed is the quantity most strongly affected by the flagellar length; beat frequency and rotational speed are largely independent of flagellar length in our measurements.   {\it Chlamydomonas} swimming speed shows sensitivity to growth media and cell cycle time, but the swimming speeds inferred by the RFT model are within the range of previous measurements of the swimming dynamics in this species \cite{MartinezRBOCJTMRJS-LLWMBWP_BiophysJ12,KhonaVRMMPVAKKDSSLBJDADSMDMJD_JBiolPhys13,FujitaTMMIMK_BiophysJ14,FolcikTHKCMRCSSNPRBRPSNN-KAP_FrontPlantSci20,BottierKTSDPB_BiophysJ19,BauerHIKWNHJKWM_iScience21}.

In conclusion, we have obtained the first time-resolved, three-dimensional images of the flagellar beat of {\it Chlamydomonas reinhardtii}.  In agreement with previous studies, we find that the beat planes of the cis and trans flagella are twisted with respect to each other, but moreover that the flagellar beat exhibits some curious asymmetries in the power and recovery stroke that are only revealed by high-speed three-dimensional imaging.  Notably, the recovery stroke of the flagella shows a distinctive chiral asymmetry between cis and trans flagella, and the power stroke is initiated with a subtle rippling or `shrugging' motion reminiscent of the beat of the uniflagellate alga {\it Euglena}, or the chirality inversion in {\it Spiroplasma} bacteria \cite{ShaevitzJLDF_Cell05}.  In the future, imaging methods capable of resolving flagellar dynamics at high frame rates and in three dimensions, such as ours and the others mentioned above, will be critical to reconciling theoretical models of flagellar beating with experimental results.

\begin{acknowledgments}
We thank B. Friedrich for helpful discussions. This work was supported by EPSRC grant EP/N014731/1 (LW) and Royal Society Exchange Grant IEC/R3/183084 (MB).
\end{acknowledgments}

\appendix

\bibliography{Master}
\end{document}